\documentclass[12pt]{article}
\usepackage{epsfig}

\def\beq{\begin{equation}}
\def\eeq#1{\label{#1}\end{equation}}
\def\eeqn{\end{equation}}

\def\beqa{\begin{eqnarray}}
\def\eeqa#1{\label{#1}\end{eqnarray}}
\def\eeqan{\end{eqnarray}}

\let\bar=\overbar

\def\Dslash{\not{\hbox{\kern-4pt $D$}}}
\def\dslash{\not{\hbox{\kern-2pt $\del$}}}

\def\msb{{\bar{\ssstyle M \kern -1pt S}}}

\def\Title#1{\begin{center} {\Large {\bf #1} } \end{center}}

\begin{document}

\Title{LHCb Results on Semileptonic B/B$_s$/$\Lambda_b$ Decays}

\bigskip\bigskip


\begin{raggedright}  

{\it Concezio Bozzi\index{Bozzi, C.}\\
Istituto Nazionale di Fisica Nucleare\\
Sezione di Ferrara\\
I-44122 Ferrara, ITALY}\\
\bigskip\bigskip
{\it on behalf of the LHCb Collaboration}\\
\bigskip\bigskip
{\it Proceedings of CKM 2012, the 7th International Workshop on the CKM Unitarity Triangle, University of Cincinnati, USA, 28 September - 2 October 2012 }
\end{raggedright}

\section{Introduction}

The LHCb experiment at the LHC has been taking pp collision data in 2011 and 2012 at $\sqrt{s}=$7 and 8 TeV respectively, integrating a luminosity in 
excess of 3~fb$^{-1}$. The large (several hundred microbarns) production 
cross sections and efficient trigger strategies allow to collect unprecedentedly large data sets for charm and beauty decays. 
As an example, about 1.2 million $\overline{\mathrm{B}}^0 \rightarrow \mathrm{D}^{*+} \mu^- {\bar{\nu}}_{\mu}$ decays~\footnote{charge-conjugation 
is always implied, unless specified otherwise} have been reconstructed 
in 1 fb$^{-1}$.  In the vast physics program which can be exploited with these large datasets, semileptonic decays play an important role.

\section{Semileptonic B Decays at LHCb}

Semileptonic B decays were firstly studied in LHCb in order to 
determine the $b\overline{b}$ production cross section~\cite{bbxsec}, the performance of flavour tagging algorithms~\cite{flavtag}, and the 
various b-hadron production fractions at the LHC~\cite{prodfrac}. More recently, semileptonic decays of  $\mathrm{B}_s$ mesons were used to test 
CP violation in neutral meson mixing, resulting in the most precise measurement of the semileptonic asymmetry in the $\mathrm{B}_s$ sector~\cite{asls}. Besides these studies, 
LHCb can contribute to the understanding of currently open issues in semileptonic decays of B, $\mathrm{B}_s$ and $\Lambda_b$ hadrons, 
including the composition of the inclusive semileptonic widths in terms of exclusive decays, measurements of form factors and determinations of 
CKM parameters $|V_{ub}|$ and $|V_{cb}|$. 

Semileptonic B decays give an experimentally clean signature, due to a charm hadron and a muon originating from a common vertex. A requirement 
on the impact parameter of the charmed hadron with respect to the primary vertex to be significantly different from zero suppresses copious background 
of charm produced promptly in the $pp$ collision. The b hadron species are determined from the reconstructed charm hadron, {\it i.e.} samples with 
D$^0$, D$^{+}$, D$_s^+$, $\Lambda_c^+$ in the final states are originating mainly from B$^-$, $\mathrm{B}^0$, $\mathrm{B}^0_s$ and $\Lambda_b^0$ 
decays, respectively. {\it Cross-feeds} due to {\it e.g.} $\mathrm{\overline{B}}^0_s \rightarrow (\mathrm{D}_s^{**} \rightarrow \mathrm{DK}) X \mu^- \nu$ or  
$\mathrm{\overline{B}}^{0,+} \rightarrow (\mathrm{D}_s \mathrm{K}) X \mu^- \nu$ (and similar baryonic decays), can be estimated by reconstructing final 
states with a D$^0$ and a charged kaon or proton, by using other available measurements and assuming {\it e.g.} isospin conservation. 
The production fractions of B$^0_s$ and $\Lambda_b$, $f_s$ and $f_{\Lambda_b}$, are measured relative to the sum of the 
production fractions of B$^0$ and B$^+$ mesons, $f_u+f_d$. Therefore, the most abundant 
B$^+$--B$^0$ cross-feed, due to $\overline{\mathrm{B}}^0 \rightarrow \mathrm{D}^{*+} \mu^- {\bar{\nu}}_{\mu}$ decays followed by 
D$^{*+} \rightarrow $D$^0 \pi^+$, is avoided. 
Relative production fractions are determined in intervals of pseudorapidity ($\eta$) and transverse momentum 
($p_T$) of the charm-muon pair, using a data sample of 3 pb$^{-1}$~\cite{prodfrac}.  
Table~\ref{tab:fracyields} shows the obtained signal and background yields. 
The results are: 
\begin{eqnarray*}
\frac{f_s}{f_u+f_d} & = & 0.134 \pm 0.004 ^{+0.011}_{-0.010}  \\
\frac{f_{\Lambda_b}}{f_u+f_d} (p_T) & = & (0.404 \pm 0.017 \pm 0.027 \pm 0.105) \times \\ 
      & &    \times [1-(0.031 \pm 0.004 \pm 0.003) p_T (\mathrm{GeV}) ] \\
\end{eqnarray*}
where units are given with $c=1$, the first and second uncertainties are respectively of statistical and systematical origin and the third one is due to the limited knowledge 
of the $\Lambda_c\rightarrow pK\pi$
branching fraction. The latter dominates the measurement of the $\Lambda_b$ production fraction, whereas systematic uncertainties due to the modeling 
of ${\mathrm{B}}_s$ semileptonic decays and $\mathrm{D}^+, \mathrm{D}_s$ branching fractions dominate the measurement of the $\mathrm{B}_s$ production fraction. 
No dependence on $p_T$ is observed in $\mathrm{B}_s$ production fraction, which is found to be in agreement with previous determinations at LEP and about 
one standard deviation lower than the Tevatron measurement. For $\Lambda_b$, the production fraction is not flat over $p_T$, in agreement with similar results 
from the Tevatron in the same $p_T$ region. A somewhat smaller fraction had been measured by the LEP experiments, on a harder $p_T$ spectrum. A detailed discussion 
and interpretation of these results is given in Ref.~\cite{HFAG2012}. 

\begin{table}[b]
\begin{center}
\begin{tabular}{lcccc}  \hline
Final state &  Signal & \multicolumn{3}{c}{Background sources:} \\  
                   &                & Prompt charm & Combinatorial & $\Lambda_c$ reflections \\ \hline
D$^0 \mu \nu X$ &  27666$\pm$ 167  & 695$\pm$43     &    1492$\pm$30 &  \\ 
D$^+ \mu \nu X$ &  9257$\pm$ 110  & 362$\pm$34     &    1150$\pm$22 &  \\ 
D$_s \mu \nu X$ &  2192$\pm$ 64  & 63$\pm$16     &   985$\pm$145 &  387$\pm$132\\ 
$\Lambda_c \mu \nu X$ &  3028$\pm$ 112  & 43$\pm$17     &   589$\pm$27 &  \\ \hline
\end{tabular}
\caption{Signal and background yields for the final states analyzed in the measurement of b-hadron production fractions.}
\label{tab:fracyields}
\end{center}
\end{table}

As a by-product of these measurements, the  D$^0$K final state was searched for P-wave D$_s$ mesons, giving the first observation of the  
semileptonic decay $\overline{\mathrm{B}}_s \rightarrow \mathrm{D}_{s2}^{*+} X \mu \overline{\nu}$ and the most precise measurement of the 
$\overline{\mathrm{B}}_s \rightarrow \mathrm{D}_{s1}^{+} X \mu \overline{\nu}$ decay.

\section{Future Prospects}

\begin{figure}[htb]
\begin{center}
\epsfig{file=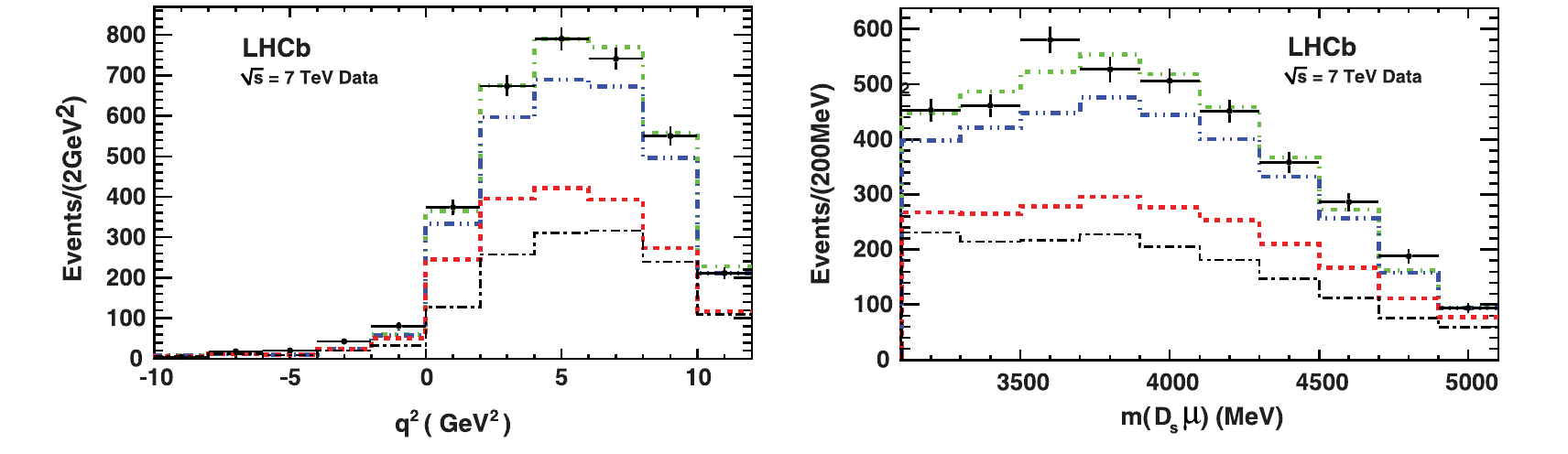,width=0.99\textwidth}
\caption{Projections of a two-dimensional fit to the $q^2$ and $m(D_s+\mu)$ distributions
of semileptonic decays including a $D_s^+$ meson. The different components
are stacked: the background is represented by a black dot-dashed line, $D_s^+$ by a red dashed
line, $D^{*+}$ by a blue dash-double dotted line and P-wave D mesons by a green dash-dotted line.}
\label{fig:dsq2}
\end{center}
\end{figure}

\begin{figure}[htb]
\begin{center}
\epsfig{file=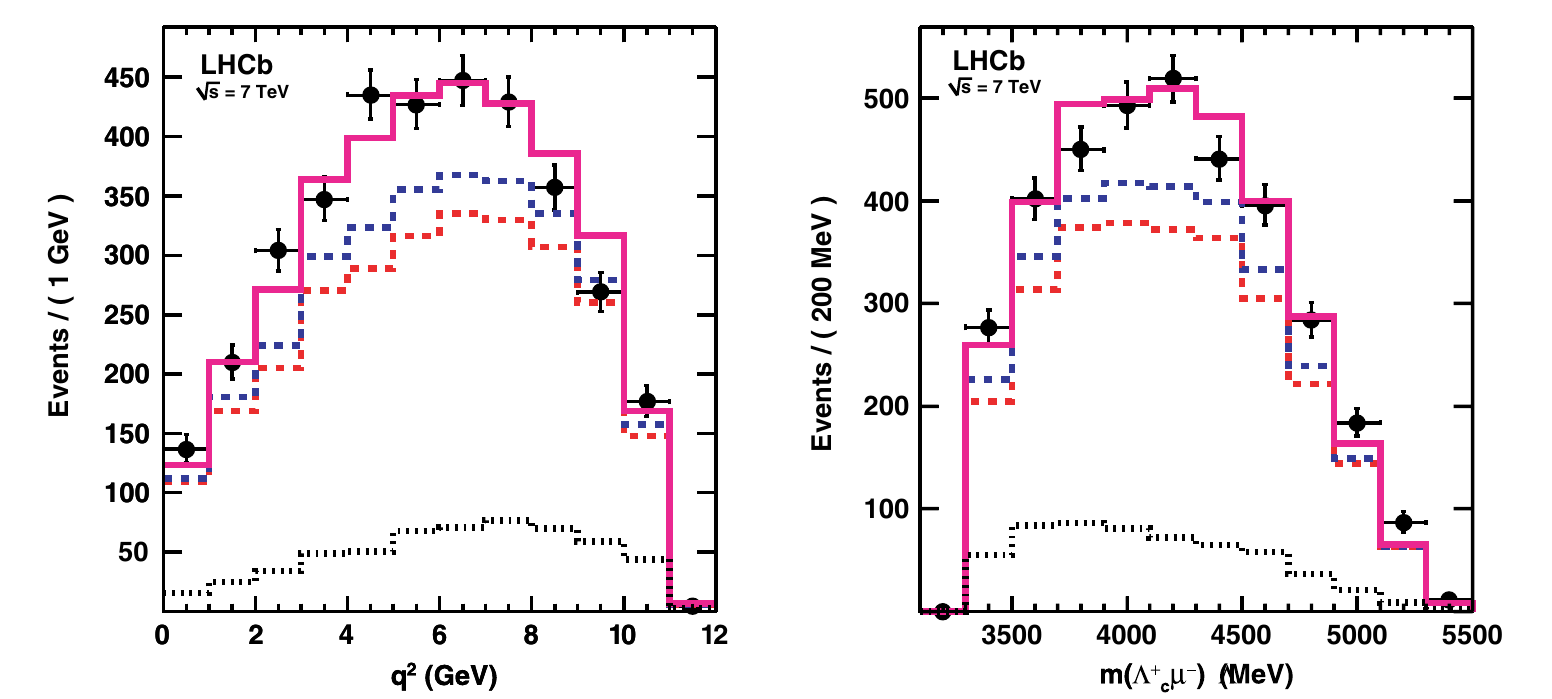,width=0.99\textwidth}
\caption{Projections of a two-dimensional fit to the $q^2$ and $m(\Lambda_c+\mu)$ distributions of semileptonic decays including a $\Lambda_c$ baryon. 
The different components are stacked: the dotted line represents the combinatoric background, the bigger dashed line (red) represents the 
$\Lambda_c^+ \mu^- \bar{\nu}$ component, the smaller dashed line (blue) the $\Lambda_c(2595)^+$, and the solid line represents the $\Lambda_c(2625)^+$ component.}
\label{fig:Lcq2}
\end{center}
\end{figure}

Measurements of the CKM matrix elements $|V_{cb}|$ and $|V_{ub}|$ in exclusive semileptonic decays imply the reconstruction,  
in the b hadron rest frame, of observables such as the squared invariant mass of the lepton pair ($q^2$). This difficult task at hadron colliders 
can be accomplished by exploiting the separation between primary and secondary vertices at LHCb. Therefore, the B flight direction vector can be 
determined and the neutrino momentum can be measured with a two-fold ambiguity. The resulting $q^2$ resolution (with a core of about 
0.4 GeV$^2$)~\cite{urquijo} is similar to that observed at the B factories. The distributions of $q^2$ and $m(\mathrm{D}_s+\mu)$ in a sample of 
$\overline{\mathrm{B}}_s \rightarrow \mathrm{D}_{s}^{+}  X \mu \overline{\nu}$ decays is shown in Figure~\ref{fig:dsq2}, where the different contributions,  
also shown, can be statistically separated~\cite{prodfrac}. Similar results have been obtained on a sample with $\Lambda_c$ baryons in the final state (see 
Figure~\ref{fig:Lcq2}). The ultimate goals of these studies will be the determination of form factors, $|V_{cb}|$ and $|V_{ub}|$ in exclusive 
B$_s$ and $\Lambda_b$ decays. 

LHCb has also good potential to measure $\overline{\mathrm{B}} \rightarrow D^{(*)} \tau^- \overline{\nu}$ decays, which recently received great 
attention after Babar measured branching fractions which exclude the Standard Model expectations at the 3.4 sigma level~\cite{bbrDtaunu}. 
However, the neutrino reconstruction method 
outlined above is much more difficult to apply in purely leptonic tau decays, due to the presence of two additional neutrinos. Therefore, three-prong decays such as 
$\tau^{\pm} \rightarrow \pi^+ \pi^- \pi^{\pm} \nu_{\tau}$ have been exploited. In this case, the neutrino reconstruction procedure can be applied and 
the tau momentum can be determined with a two-fold ambiguity, which becomes four-fold on the B meson momentum. Moreover, care must be taken in order 
to avoid non-physical solutions due to the finite momentum and vertex resolutions. A recent measurement of 
$\mathrm{B}_{(s)} \rightarrow \mathrm{D}_{(s)} \pi \pi \pi$ decays~\cite{BDpipipi} indicates that similar decays with high track multiplicity can be 
reconstructed in LHCb, with low combinatorial background. 

The long standing puzzle of the composition of the inclusive semileptonic rate in terms of the exclusive decays can also be investigated, by 
reconstructing decays in P-wave charm mesons as well as in radial excitations. 
An investigation of three-body decays of these particles, already in progress in hadronic B decays
~\cite{BDpipipi}, will also help in normalizing their corresponding semileptonic B branching fractions, 
which are typically computed by assuming that P-wave charm mesons decay in two-body final states only.

\section{Conclusions}

In conclusion, semileptonic B/B$_s$/$\Lambda_b$ decays are an important part of the LHCb physics program. They have been succesfully used to measure the  
$\mathrm{b\bar{b}}$ cross section and b hadron production fractions at the LHC, and to establish the most accurate limit on CP violation in the mixing of 
B$_s$ mesons. Studies of B$_s$ and $\Lambda_b$ decays have been performed, which will eventually lead to novel determinations of the $|V_{cb}|$ and $|V_{ub}|$  
CKM matrix elements, which can help to clarify the existing tension between measurements with inclusive and exclusive semileptonic B decays. 
A measurement of $\overline{\mathrm{B}} \rightarrow D^{(*)} \tau^- \overline{\nu}$ decays seems also feasible at LHCb, although neutrino reconstruction will be challenging. 
Precise measurements of semileptonic decays in higher D meson excitations will allow to reduce systematic errors on other measurements, 
and possibly solve long-standing puzzles in semileptonic B decays.

\end{document}